\documentclass[12pt]{iopart}
\usepackage{iopams}
\usepackage{graphicx}
\usepackage{bm}

\newcommand{\Z}{\mathbb{Z}}
\newcommand{\N}{\mathbb{N}}
\newcommand{\beq}{\begin{equation}}
\newcommand{\eeq}{\end{equation}}

\newcommand{\n}{\mathbf{n}}

\newcommand{\x}{\mathbf{x}}
\newcommand{\eps}{\varepsilon}
\newcommand{\mafrac}{\sqrt{\frac{8 \pi}{3}}}

\begin{document}
\title{The fitting problem in a lattice Universe}

\author{Julien Larena$^{1}$}

\address{$^1$ Department of Mathematics, Rhodes University, Grahamstown, 6140, South Africa}

\ead{j.larena@ru.ac.za}

\begin{abstract}
We present a regular cubic lattice solution to Einstein field equations that is exact at second order in a small parameter. We show that this solution is kinematically equivalent to the Friedmann-Lema\^itre-Robertson-Walker (FLRW) solution with the same averaged energy density. This allows us to discuss the fitting problem in that framework: are observables along the past lightcone of observers equivalent to those in the analogue FLRW model obtained by smoothing spatially the distribution of matter? We find a criterion on the compacity of the objects that must be satisfied in order for the answer to this question to be positive and given by perturbative arguments. If this criterion is not met, the answer to this question must be addressed fully non perturbatively along the past lightcone, even though the spacetime geometry can be described perturbatively.

\end{abstract}

\section{What is the fitting problem?}

Cosmology is unique among physical sciences for a certain number of reasons. First, the Universe is given once and for all, and there is no possibility to compare it to another Universe. This can usually be overcome by supposing that the initial conditions for the Big Bang model must be generic in some reasonable sense, or that some mechanism (e.g. inflation) is responsible for making them generic. Second, we are the only available observers in the Universe; there might be other observers, but we do not have access to their observations. The only piece of information we have on the Universe comes from our past lightcone, and a few local (geological) measurements on our worldline. Therefore, in general, one cannot rely purely on observations to fully determine the nature and dynamics of the Universe: one has to introduce extra assumptions on the theory of gravitation, the geometry of the Universe on large scales and the physical nature of its matter content. In the present paper, we will suppose throughout that gravity obeys the laws of General Relativity, and we will concentrate on the other two points: the geometry of the Universe and its matter content. In the standard model of cosmology, it is assumed that 'on average', on sufficiently large scales, the distribution of matter in the Universe is well described by a set of perfect fluids whose energy densities and pressures are locally homogeneous and isotropic; this results in Friedmann-Lema\^itre-Robertson-Walker (FLRW) Universes with spatial sections that are maximally symmetric, i.e. determined entirely by their constant Gaussian curvature. This assumption is based on the observed almost isotropy of the Cosmic Microwave Background around us, together with the Copernican Principle, and is usually called the Cosmological Principle. It is clear that it is an extrapolation outside our past lightcone, since the notion of average implicitly present in this principle tells us something about the spatial distribution of matter, starting from its distribution along our past lightcone. In a nutshell, the fitting problem \cite{Ell84} can be summarized by the question: does the effective FLRW model obtained by extrapolating the observed properties down our past lightcone coincide with the effective FLRW model obtained by smoothing the spatial distribution of matter? Of course, this question is not independent on the set of observers used to define the notion of spatial distribution: it makes use of a preferred set of observers, called fundamental in the standard model; usually, in the late-time Universe, the fundamental observers are supposed to be comoving with virialised objects such as galaxies, so as to include us among fundamental observers. In this paper, we will try and address the fitting problem by considering a special dynamical solution to the field equations consisting in a regular cubic lattice of initial cell size $L$ and of objects of equal masses $M$. The solution is exact at second order in the small parameter $\sqrt{M/L}$ and we will see that it exhibits, on average, the same dynamical behaviour than the equivalent flat FLRW model with a non-relativistic fluid of density $\rho=M/L^{3}$, therefore showing that, at second order in $\sqrt{M/L}$, there is no backreaction in the model \cite{Bruneton:2012cg}. Then, because the solution for the metric is exact at this order, we will be able to solve the Sachs equations at second order in $\sqrt{M/L}$ in order to reconstruct observables such that the distance-redshift relation. We will see that this solution for observables presents some divergences linked with the compacity of the object: if the extension $\eta$ of the objects is too small, the perturbative expansion of the solution of the Sachs equations is no longer valid, even though the perturbative expansion of the solution of the Einstein field equations remains stable. Namely, we will show that observables in this model remain very close to the observables calculated in the analogue FLRW model with energy density $\rho=M/L^{3}$, as long as the parameters of the lattice obey: $\frac{M}{L}\ll \mathcal{O}\left(1\right)\left(\frac{\eta}{L}\right)^{4}$. If this condition is not satisfied, then observables cannot be calculated perturbatively, even though the metric is well approximated by the perturbative expansion, and one must solve the full system of Sachs equations \cite{Bruneton:2012ru}. Results regarding this complete integration are hinted at in this paper. These results illustrate the importance of the fitting problem in cosmology: the kinematically averaged FLRW model and the FLRW reconstructed by fitting observations might differ significantly (if calculated perturbatively, at least), even in the absence of (kinematical) backreaction. We work in units $G=c=1$.

\section{A lattice Universe: kinematics and observables}

\subsection{The cubic lattice solution}

Let us start by describing the lattice solution (see \cite{Bruneton:2012cg} for a complete derivation); we will only sketch the results and discuss their implications. We start with a cubic lattice of size $L$ with identical masses $M$ at the centre of each cell. If the masses on the lattice are to represent typical galaxies, we can choose, as our typical parameters $M\sim 10^{11}M_{\odot}$ and $L\sim 1\mbox{ Mpc}$, where $M_{\odot}\sim 10^{30}$ kg is the Solar mass, and $L$ is of order of the intergalactic distances. Then, the natural parameter of the lattice is $R_{S}/L\sim 10^{-8}\ll 1$, where $R_{S}=2M$ is the Schwarzschild radius of the masses. Therefore, we can look for a solution expanded into powers of $\sqrt{M/L}$ (in \cite{Bruneton:2012cg} we prove that there is no perturbative solution in powers of $M/L$); this will lead to linearised field equations that can be solved exactly. We choose coordinates that are comoving with the masses: $g_{00}=-1$, and spatial coordinates are Cartesian and adapted to the symmetries of the lattice. The distribution of matter is therefore a three dimensional Dirac comb with the masses located at $\x_{\n}=L\mathbf{n}$, $\n\in\Z^{3}$; the energy momentum tensor is then: $T_{ab}=T_{00}\delta^{0}_{a}\delta^{0}_{b}$ such that:
$$
T_{00} \propto M \sum_{\n \in \Z^3} \delta^{(3)}(\x-L\n ).
$$
Actually, the field equations do not have a solution for such a source term \cite{Korotkin:1994dw,Bruneton:2012cg}, because the formal series solution presents a UV divergence coming from the point-like nature of the masses. Therefore, we introduce a UV cut-off by giving a small but finite extension to the masses, $\eta$ and by replacing the Dirac deltas by their standard approximation: $\delta(x-nL)\sim \frac{1}{\eta\sqrt{\pi}}e^{-\frac{(x-nL)^{2}}{\eta^{2}}}$. Then, we write the source term in Fourier series, and  we expand the field equations in powers of $\sqrt{M/L}$ and solve them order by order, to find the following solution at second order: $\forall i\in\{1,2,3\}, g_{0i}=0$, and $\forall (i,j)\in\{1,2,3\}^{2}$:
\begin{eqnarray}
\label{mainresultmetric}
g_{ij}&=& \delta_{ij} \left[1+ 2\eps\sqrt{\frac{M}{L}}  \mafrac \frac{t}{L} +\frac{2M}{L}\left(f_{\eta}(\mathbf{x})+ \frac{2 \pi t^2}{3 L^2} \right) \right]\nonumber\\
 & &+\frac{M}{L} t^2\partial_{ij}^2 f_{\eta}(\mathbf{x}) 
\end{eqnarray}
where $\epsilon=\pm 1$ and:
\begin{equation}
f_{\eta}(\x)=\frac{1}{\pi}\sum_{\mathbf{n}\in\Z^{3}_{*}}\frac{e^{-\frac{\pi^{2}|\mathbf{n}|^{2}\eta^{2}}{L^{2}}}}{|\mathbf{n}|^{2}}e^{\frac{2\pi}{L}i\mathbf{n}.\x}.
\end{equation}
Let us insist on the fact that this solution is exact at order $M/L$. We can now calculate the rate of expansion between two masses of the lattice. For that, consider two  masses on the x-axis (all the other axes are equivalent, by symmetry), separated by a coordinate distance $NL$, for $N$ an integer. The physical distance between the two masses is given by $l(t)=\int_{0}^{NL}\sqrt{g_{xx}}dx$, and, expanding the square root to order $M/L$ we find the effective scale factor of the lattice:
\begin{equation}
a(t) \equiv \frac{l(t)}{NL}=1+\eps \mafrac \sqrt{\frac{M}{L^3}} t - \frac{2 \pi M t^2}{3 L^3}.
\end{equation}
The Hubble flow defined by $H(t) = \dot{a}(t)/a(t)$ is then found to be, at order $M/L$: $H(t) = \eps \mafrac \sqrt{\frac{M}{L^3}} - \frac{4 \pi M t}{L^3}$. Therefore, defining the initial Hubble rate $H_0 = \eps \mafrac \sqrt{\frac{M}{L^3}}$ and choosing the expanding solution, $\eps=1$, we get:
\beq
\label{Hfinal}
H(t) =  H_0 -\frac{3}{2} H_0^2 t +\mathcal{O}(H_0^3)
\eeq
and this corresponds exactly, at order $M/L\propto H_{0}^{2}$ to a flat FLRW model filled with non-relativistic dust. The result is actually valid at order $\left(M/L\right)^{3/2}$ \cite{Bruneton:2012cg}. Thus, the model with discrete masses on a cubic lattice, once smoothed, is identical to a FLRW model with dust, with the corresponding smeared energy density. This means that, from purely kinematical considerations, one cannot distinguish between the average, homogeneous fluid description of the lattice and the exact behaviour of this lattice: there is no backreaction (in the sense of \cite{Buchert:1999er}) associated with spatially smoothing the lattice.

\subsection{Observables, compacity and the fitting problem}

Now that we have a solution of the field equations that does not display backreaction, we can try and address the fitting problem by comparing observables in the lattice with observables in the corresponding, smoothed FLRW Universe with the same kinematics: any discrepancy between the two will be a sign that there exists a fitting problem. In order to carry the comparison, we calculate the distance/redshift relation in the lattice. The $4$-velocity of an observer in one of the objects of mass $M$ is given by $u^{a}=(1,0,0,0)$ according to our choice of coordinates, and we define $\lambda$, an affine parameter down light rays. We denote by $O$ and $S$ the locations of observer and the source respectively. Given the normalisation chosen in \cite{Bruneton:2012ru}, the null vector of a past-directed light ray, $k^{a}$ is such that $k^{0}_{O}=1$, so that we have, for the redshift: $1+z(\lambda)=\frac{\left(k^{a}u_{a}\right)_{S}}{\left(k^{a}u_{a}\right)_O}=k^{0}\left(\lambda\right)$. Therefore, we can solve the $0$ component of the null geodesic equation order by order in terms of $\sqrt{M/L}$. The distance is obtained similarly by solving Sachs equations \cite{Sachs:1961zz} expanded at order $M/L$. The Sachs equations are actually exactly solvable in this perturbative scheme because the equations for the isotropic expansion and the shear decouple from each other at that order. Details of the calculations can be found in \cite{Bruneton:2012ru}. In terms of the past-directed affine parameter $\lambda<0$ (and $\lambda=0$ at the observer), we find that, at order $M/L$:
\begin{eqnarray}
z(\lambda)&=& -\sqrt{\frac{M}{L}}\mafrac \frac{\lambda}{L} \nonumber \\ 
& &+\frac{M}{L}\left(\frac{14 \pi \lambda^2 }{3 L^2}+\left[ f_{\eta}(\x(\lambda))-\lambda \partial_{i} f_{\eta}(\x(\lambda)) v^i\right]^{\lambda}_{0}\right)\\
r_A(\lambda) &=& -\lambda + \frac{2 \pi}{3}\frac{M}{L}\frac{\lambda^3}{L^2} 
\left[1 +\sum_{(n,p,q)\in\mathcal{D}_{\mathbf{v}}} e^{-\frac{\pi^{2}(n^2+p^2+q^2)\eta^{2}}{L^{2}}}\right]\nonumber \\
&+&\frac{2}{\pi}\frac{M}{L} \sum_{\mathbf{n}\in \N_{*}^3 \setminus \mathcal{D}_{\mathbf{v}}}  e^{-\frac{\pi^{2} (n^2+p^2+q^2)\eta^{2}}{L^{2}}} \times \nonumber \\
& &\sum_{l=1}^{l=4}  \left[ -\lambda  \frac{\cos\left(\frac{2 \pi \lambda \mathbf{v}.\mathbf{u}_l }{L}\right)}{(\mathbf{v}.\mathbf{u}_l)^2}  +\frac{L}{\pi} \frac{\sin\left(\frac{2 \pi \lambda \mathbf{v}.\mathbf{u}_l }{L}\right)}{(\mathbf{v}.\mathbf{u}_l)^3} \right].
\end{eqnarray} 

\begin{center}
\begin{figure}
\includegraphics[width=\textwidth]{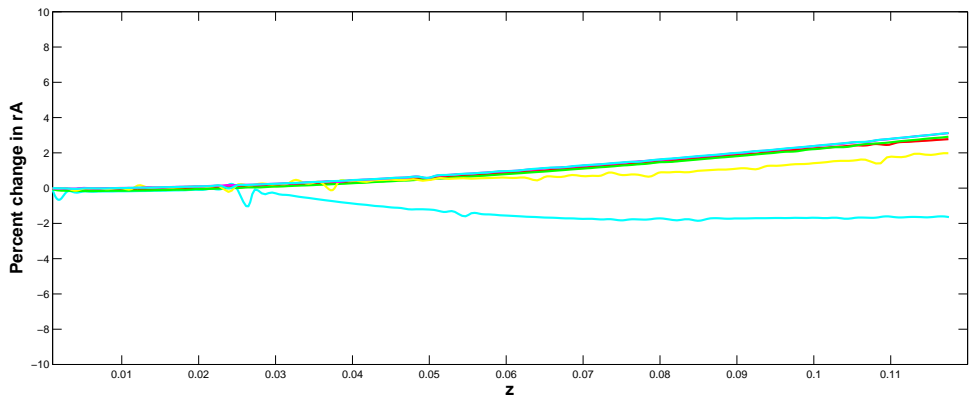}
\caption{\label{fig1}Percent change $\delta r_{A}(z)$ in the angular distance between a lattice with $M=10^{12}M_{\odot}$, $L=1$ Mpc, $\eta=0.01 L$ and its FLRW counterpart, obtained from a complete integration of Sachs equations. The result is presented for 25 different directions on the celestial sphere of the observer located at the centre of one of the masses.}
\end{figure}
\end{center}

Here: $\mathbf{u}_1=(n,p,q), \, \, \mathbf{u}_2=(n,-p,-q), \, \, \mathbf{u}_3=(n,p,-q),\, \, \mathbf{u}_4=(n,-p,q)$, and $\mathcal{D}_{\mathbf{v}}=\{(n,p,q) \in\N_*^3 : \exists\, l\in\{1,2,3,4\} / \mathbf{u}_l.\mathbf{v}=0\}$. This means that the first sum is over all the triplets that cancel one at least of the $\mathbf{u}_{l}.\mathbf{v}$, whereas the second sum is over all the other triplets. These expressions coincide with their FLRW counterparts for the FLRW model obtained by smoothing the distribution of masses of the lattice, up to {\it a priori} small corrections proportional to $M/L$ (the parts that are non-polynomial in $\lambda$). Actually, it turns out that the additional terms in the expression for $r_{A}\left(\lambda\right)$ are not generally small, because some denominators in the second sum become extremely small and the corrections to the FLRW distance become of order $\sqrt{M/L}$, or even $1$, instead of being of order $M/L$; see \cite{Bruneton:2012ru} for a detailed discussion of these effects. By carefully studying these additional terms, we arrive at the conclusion that the perturbative corrections to the FLRW distance/redshift relation remain small provided:
\begin{equation}
\label{mainbound}
\frac{M}{L} \ll \mathcal{O}(1) \times \left(\frac{\eta}{L}\right)^4.
\end{equation}
This relation between the mass of the object, $M/L$, and the compacity of the lattice, $\eta/L$, shows that if objects are too compact, the perturbative expansion breaks down, as far as the calculation of observables down a past lightcone is concerned, even though the perturbative calculations remain a good way of estimating the spacetime geometry (i.e. of solving the Einstein field equations). Similar problems were encountered in perturbations of an FLRW background in \cite{Clarkson:2011uk}.
If this criterion is not satisfied, the perturbative calculations cannot be trusted, as second order terms (in $\sqrt{M/L}$) become of order $\mathcal{O}(1)$: one needs to integrate the system of Sachs equations without any perturbative expansion, thus retaining the coupling between isotropic expansion and shear. This is an ongoing work \cite{Larena2012} and preliminary results indicate that when Eq.~(\ref{mainbound}) is not satisfied, the contribution of the shear modifies significantly the FLRW observables: the corrections are usually smaller than the order $1$ corrections predicted by the perturbative expansion presented here, but they are definitely significant to raise the issue of a fitting problem. For example, Fig.~\ref{fig1} shows the percent change $\delta r_{A}(z)=100\times\left(r_{A}(z)-r_{A}^{FLRW}(z)\right)/r_{A}^{FLRW}(z)$ in the angular distance between a lattice with $M=10^{12}M_{\odot}$, $L=1$ Mpc, $\eta=0.01 L$ (lattice of galaxy-like objects) that does not satisfy the criterion (\ref{mainbound}) and the equivalent smoothed FLRW model, obtained from a complete integration of Sachs equations. We see that the 'divergence' problem encountered for such lattices when using perturbative methods is somehow 'cured' by solving the full system, even though, differences appear and seem to be systematically increasing with the redshift, irrespective of the direction on the sky.

Table~\ref{Tab1} presents the typical values of $M$, $L$ and $\eta$ for some lattices of standard astrophysical objects and shows whether the criterion (\ref{mainbound}) is satisfied or not for such lattices.

\begin{table}
\begin{center}
\begin{tabular}{|c|c|c|c|c|}
\hline
Object & \begin{tabular}{c}$R_{S}$\\ (Mpc)\end{tabular} & $\begin{tabular}{c}L\\ (Mpc)\end{tabular}$ & \begin{tabular}{c}$\eta$\\ (Mpc)\end{tabular} & \begin{tabular}{c}
Criterion\\
passed\end{tabular} \\
\hline
\begin{tabular}{c}Neutron\\
star\end{tabular} & $10^{-19}$ & $10^{-6}$ & $10^{-18}$  & No\\
\hline
\begin{tabular}{c}
Galaxy\\
 (disk)
 \end{tabular} & $10^{-8}$ & $1$ & $10^{-2}$ & \begin{tabular}{c} No\\
(Marginally)\end{tabular} \\
\hline
\begin{tabular}{c}
Galaxy\\
(disk+DM halo)
\end{tabular} & $10^{-7}$ & $1$ & $0.05$ & \begin{tabular}{c}Yes\\ (Marginally)\end{tabular}\\
\hline
\begin{tabular}{c}
Galaxy\\
cluster\end{tabular} & $10^{-4}$ & $30$ & $20$ & Yes\\
\hline
\end{tabular}

\end{center}
\caption{\label{Tab1}Some typical lattices and their characteristic parameters. The choices are only indicative. The last column answers the question: does such a lattice satisfy the criterion (\ref{mainbound})?}
\end{table}

We see that a lattice of galaxies composed of their disk only marginally fails to pass the criterion, whereas when we include the Dark Matter halo, they pass the test marginally. A lattice of cluster-like objects passes the criterion easily, but $\eta/L\sim 1$, and one can hardly talk of a lattice of separated objects in that case: such objects could not really be considered as virialised, independent objects as we did in this work.

\section{Discussion}

We have presented a toy model of the Universe in the form of a regular cubic lattice of equal masses of typical size $\eta$ whose kinematics is identical on large scales to the FLRW model obtained by smoothing the distribution of masses; this model does not exhibits any backreaction. We have seen that, despite the fact that a perturbative expansion in terms of $R_{S}/L$ gives a very good approximation for the geometry of space-time, it is not suitable for the accurate calculation of observables in the model if the objects are too compact. Specifically, we have shown that a perturbative calculation of observables can be trusted only if the parameters of the lattice satisfy (\ref{mainbound}). If this is not the case, a non perturbative approach is needed to fully take into account the effect of the Weyl curvature sourcing the shear of bundles of null geodesics. Somehow, this was to be expected: the perturbative calculation decouple the shear from the isotropic expansion, making the observables independent on the Weyl curvature, but we know that in a mostly empty Universe (masses very compact), the behaviour of null ray bundles must be dominated by the Weyl curvature. The bound (\ref{mainbound}) gives a quantitative criterion to decide what 'too compact' means. This also illustrates the importance of the fitting problem: if (\ref{mainbound}) is not satisfied, the FLRW reconstructed by smoothing spatially the kinematics of the model differs systematically and sometimes significantly from the FLRW model fitting the observations on the past lightcone of observers. Choosing the order of magnitude of the parameters of the model to represent something 'realistic' is difficult, but it is interesting to note from Table~\ref{Tab1} that galaxy-like objects and lattices are exactly at the transition between the lattices that pass the criterion (\ref{mainbound}) and those which do not. This might be extremely important in the precise characterisation of the properties of Dark Energy, and the detailed exploration of the consequences of this bound as well as the precise non perturbative estimates in the cases when it is not satisfied are the subject of an ongoing work \cite{Larena2012}.

\section*{References}
\bibliographystyle{unsrt}
\bibliography{Torus_biblio}

\end{document}